\documentclass[prb,aps,twocolumn, amsfonts,amsmath,amssymb,nofootinbib,superscriptaddress]{revtex4-2}
\pdfoutput=1

\usepackage{amsmath}
\usepackage[dvipsnames]{xcolor}
\usepackage{hyperref}
\hypersetup{
	colorlinks=true,
	linkcolor=blue,   
	urlcolor=blue
}
\usepackage{graphicx}
\usepackage{tikz}

\newcommand{\op}[1]{\mathcal{O}^{(#1)}}
\newcommand{\nn}{\nonumber\\}
\newcommand{\goesto}{\rightarrow}
\renewcommand{\max}{\text{max}}
\newcommand{\eff}{\text{eff}}
\newcommand{\cl}{\text{cl}}
\newcommand{\q}{\text{q}}
\newcommand{\half}{\frac{1}{2}}
\newcommand{\mmax}[1]{\mathcal{M}^{#1}_\max}

\newcommand\be            {\begin{equation}}
\newcommand\ee            {\end{equation}}
\newcommand\ba            {\begin{aligned}}
\newcommand\ea            {\end{aligned}}
\newcommand{\p}{\partial}
\renewcommand{\t}{\theta}
\newcommand{\mcm}{\mathcal{M}}

\begin{document}
	
\title{Spontaneous breaking of multipole symmetries}
\author{Charles Stahl}
\affiliation{Department of Physics and Center for Theory of Quantum Matter, University of Colorado Boulder, Boulder CO 80309, USA}
\author{Ethan Lake}
\affiliation{Department of Physics, Massachusetts Institute of Technology, Cambridge, MA, 02139}
\author{Rahul Nandkishore}
\affiliation{Department of Physics and Center for Theory of Quantum Matter, University of Colorado Boulder, Boulder CO 80309, USA}

\begin{abstract}
Multipole symmetries are of interest both as a window on fracton physics and as a crucial ingredient in realizing new universality classes for quantum dynamics. Here we address the question of whether and when multipole symmetries can be spontaneously broken, both in thermal equilibrium and at zero temperature. We derive generalized Mermin-Wagner arguments for the total or partial breaking of multipolar symmetry groups and generalized Imry-Ma arguments for the robustness of such multipolar symmetry breaking to disorder. We present both general results and explicit examples. Our results should be directly applicable to quantum dynamics with multipolar symmetries and also provide a useful stepping stone to understanding the robustness of fracton phases to thermal fluctuations, quantum fluctuations, and disorder. 
\end{abstract}
	
\date{\today}
	
\maketitle

\section{Introduction}

Hamiltonians invariant under polynomial symmetry transformations conserve not only charge, but also various multipole moments of charge. Such `multipolar' symmetries are known to offer a robust route to ergodicity breaking \cite{PPN, KHN, Sala, Moudgalya, SLIOM, commutant}, and also to exotic universality classes of quantum dynamics \cite{Iaconis1, GLN, nonabelian, glorioso, MKH, Feldmeier, Iaconis2}. They are known to arise in `fracton' phases of quantum matter \cite{Chamon, Haah, VHF1, VHF2, NH, PretkoRadzihovsky}, the key dynamical properties of which are known to descend from conservation laws on multipole moments of charge \cite{Pretko1, BB,  Gromov2019}. They are also known to arise (in a prethermal sense) in various ultracold atom platforms \cite{KHN, Bakr, Aidelsburger}. There are thus multiple reasons for thinking about systems with multipolar symmetries. However, just because a symmetry is present in the {\it Hamiltonian} does not mean that it will be present in the {\it state}; there is always the possibility of spontaneous symmetry breaking (SSB). 

For conventional symmetries, there exist general theorems which constrain the settings in which SSB can occur. In clean systems, the relevant theorem is due to Mermin and Wagner \cite{MerminWagner}, and involves the physics of (thermal or quantum) fluctuations of the Goldstone modes associated with SSB, whereas in disordered systems the key results are due to Imry and Ma \cite{ImryMa, Vojta2013}, and also Aizenman and Wehr \cite{Aizenman}, and involve the physics of order parameter deformation for local alignment with disorder. Multipolar symmetries, however, allow for a much richer pattern of possible symmetry breakings (including breaking some but not all of the multipolar symmetries), and the analogous theorems have not yet been derived, except in the special case of isotropic clean systems with total breaking of the symmetry \cite{Griffin2013Multi}.

In this work, we place generalized Mermin-Wagner and Imry-Ma constraints on the total and/or partial breaking of multipolar symmetries, in both clean and disordered systems. Along the way we also discuss the exotic Goldstone modes associated with total or partial SSB of multipolar symmetries. We will also provide explicit models of multipole symmetry breaking, to give intuition for these unusual forms of SSB. 

Throughout, we consider only multipole groups where the underlying internal group is continuous and abelian. For concreteness, we will say it is $U(1)$. Multipole groups with a nonabelian underlying symmetry suffer a cascade effect where the dynamics in at least one direction must be trivial~\cite{nonabelian} and we shall not discuss them here. We note that specific examples of spontaneous symmetry breaking of multipolar symmetries have been discussed (sometimes in a dual language) in \cite{elastic1, elastic2, elastic3, elastic4, elastic5, FS1, FS2}. Our goal here is to place general constraints on when certain symmetry breaking phase transitions involving multipolar symmetries can occur. 

Many of the differences from the ordinary Mermin-Wagner theorem come from the soft dispersions of the relevant Goldstone modes. Although soft Goldstone modes exist in other models, they generically only emerge at fine-tuned points. As we will see, the multipole symmetries make the soft modes natural, in the sense that the theories do not have to be fine-tuned. In addition, in Sec.~\ref{sub:partial} we will see that there are exotic partial symmetry-breaking options that are not available in the presence of non-multipolar soft modes. 

This paper is organized as follows. In Section \ref{multipolegroup} we introduce the multipole group, and how to build field theories invariant under it. In section \ref{sec:full_breaking} we discuss generalized Mermin-Wagner arguments for situations in which a multipole group is spontaneously broken down to the trivial group, while in Section~\ref{sec:partial} we explore the more subtle case where a nontrivial subgroup is preserved. This analysis allows us to construct examples where a single symmetry-breaking pattern can be described by several distinct types of Goldstone modes, as well as an example where a continuous symmetry is spontaneously broken in one dimension at $T=0$. 
We then discuss generalized Imry-Ma arguments in the presence of quenched disorder in Section~\ref{sec:disord}. Finally, in Section~\ref{sec:example} we consider an explicit lattice model that illustrates some of the ideas of the previous sections. We conclude with a discussion of open questions in \ref{sec:disc}.

\section{The multipole group}
\label{multipolegroup}

The multipole group is well-explained in Ref.~\cite{Gromov2019}. For concreteness, and since we will be interested in situations where a multipolar symmetry group is spontaneously broken, we will describe the multipole group in terms of its action on a compact scalar field $\phi$, which will play the role of a Goldstone boson for the broken multipolar symmetry. We will always imagine $\phi$ as constituting the phase mode for some microscopic order parameter $\psi \sim e^{i\phi}$, with $\psi$ transforming by various $U(1)$ phases under the multipole group. 

The multipole group generalizes the internal shift symmetry $\phi(x) \mapsto \phi (x) +c$ by allowing a shift by some set of polynomials involving the spatial coordinates, viz. $\phi (x) \mapsto \phi (x) + \lambda_\alpha P^\alpha(x)$. The variables $\lambda_\alpha$ are symmetry parameters, while $\alpha$ labels the set of polynomials $P^\alpha(x)$. These so-called polynomial shift symmetries~\cite{Griffin2015} all commute with each other. It is helpful to limit ourselves to homogeneous polynomials. We can label these as $P_a^{\alpha}$, where $a$ is the degree of the polynomial and $\alpha$ is now an index that runs over the homogeneous of degree $a$.

The full structure of the multipole group comes into play when we also include spatial symmetries. For example,  translation in the $x_1$ direction will fail to commute with any polynomial shift where the polynomial is a function of $x_1$. Thus, if we want to consider a collection of polynomial shift symmetries, we must consider whether that collection closes under conjugation by translations and rotations. If it does not, we must either exclude the offending translations or rotations, or expand the set of polynomial shift symmetries. The result is a multipole symmetry group~\cite{Gromov2019}.

\subsection{Examples} \label{sub:examples}

Reference~\cite{Gromov2019} includes discussion of some multipole groups; we will review a few here. The simplest case is the maximal multipole group $\mathcal{M}^a_\text{max}$, which includes all shifts by polynomials of degree $a$ or less. Individual polynomials can be written as
\begin{align}
P_c^{\gamma} = \mu^{\gamma}_{I_c}x^{I_c}, \label{eqn:basis}
\end{align}
where $\gamma$ is an index that runs over polynomials of degree $c$, and $c\le a$. The composite index $I_c = \{i_1,\dots,i_c\}$  expands as $x^{I_c}=x^{i_1}\cdots x^{i_c}$, and similarly for $\partial_{I_c}$. Each  matrix  $\mu^{\gamma}$ is fully symmetric on its $I_c$ indices. This group also includes all translations and rotations. We will use $\mmax{-1}$ to denote the rotations and translations without any internal symmetry.

An example of a multipole group that contains all translations and rotations but is \emph{not} the maximal multipole group is the group generated by shifts of the form
\begin{align}
\phi \goesto \phi + \lambda_0 P_0^0 + \lambda_{i} P^{i}_1 + \lambda_{\beta} P^{\beta}_2,
\end{align}
where the degree-0 polynomial is $P_0^0=1$. The other polynomials are
\begin{align}
P_1^{i} = x^i,\quad \quad P_2^{\beta} &= \mu^{\beta}_{ij} x^i x^j,
\end{align}
where $\mu^{\beta}$ are a basis for the traceless symmetric $d\times d$ matrices. Let us call this group $\mathcal{M}^2_{\text{sym}}$. Recall that the maximal quadrupole group $\mathcal{M}^2_{\text{max}}$ is already only built from symmetric matrices $\mu$. The tracelessness condition thus only removes one polynomial from the set. 

This set of polynomial shift symmetries is compatible with all rotations because no rotation will generate a traceful matrix from a traceless one. The set of symmetric matrices, seen as a representation of the group of rotations, decomposes into two independent irreducible representations. One of these is the set of traceless matrices while the other is the single matrix $\delta_{ij}$. In fact, the set of polynomial shift symmetries consisting of constant and linear shifts along with shifts of the form $\delta \phi \propto x^ix^i$ is also compatible with all rotations. We could call this group $\mathcal{M}^2_\text{tr}$.

There is one multipole group worth mentioning that does not include all rotations. This is the multipole group corresponding to Haah's U(1) code~\cite{Haah2017, BB, Gromov2019}. We will explain the correspondence in the next subsection. The group itself consists of all translations, a single rotation about the $(1,1,1)$ axis (on the cubic lattice), and shifts by five polynomials~\cite{Gromov2019}. These are
\begin{align}
P_0^0 &= 1, \nn
P_1^1 &= x_1 - x_2,\nn
P_1^2 &= x_1 + x_2 - 2x_3\nn
P_2^1 &= (x_1 - x_2) (x_1 + x_2 - 2x_3)\nn
P_2^2 &= (2x_1 - x_2 - x_3) (x_2 - x_3).
\end{align}
Although this looks complicated, we can simplify the presentation by first choosing a new spatial basis and then also new basis for the polynomials.

If we define new variables $x = (x_1 - x_2)/\sqrt{2}$, $y = (x_1 + x_2 - 2x_3)/\sqrt{6}$, and $z = (x_1 + x_2 + x_3)/\sqrt{3}$, then we can write the Haah group as all translations, a single rotation in the $x-y$ plane, and
\begin{align}
P_1^1 &= \sqrt{2} x , \nn
P_1^2 &= \sqrt{6} y, \nn
P_2^1 &= \sqrt{12} xy\nn
P_2^2 &= 6y^2 +2\sqrt{12} xy - 6x^2.
\end{align}
Finally, a basis change and redefinition for the polynomials allows us to write
\begin{align}
P_1^1 &= x , \nn
P_1^2 &= y, \nn
P_2^1 &= xy\nn
P_2^2 &= y^2 - x^2.
\end{align}
Note that no polynomials depend on $z$.

The Haah multipole group thus has a product structure, $\mathcal{M}_\text{Haah} = \mathbf{R} \times \mathcal{M}^2_\text{sym}$, where the $\mathbf{R}$ corresponds to translations along $z$. The second group $\mathcal{M}^2_\text{sym}$ is the submaximal quadrupole group in 2 dimensions, only containing quadrupole shifts corresponding to symmetric traceless tensors.

\subsection{Multipole field theories} \label{sub:field}

We can now consider building IR field theories for phases in which a multipolar symmetry is completely spontaneously broken, with no residual unbroken subgroup. Reference~\cite{Gromov2019} describes the process in detail, so in the following we will be somewhat succinct. 
We will continue to write things down in terms of the compact scalar $\phi$, which (at least in the cases for which SSB is not preempted by strong fluctuations) is to be viewed as the Goldstone for the sponatneously broken symmetry. 

Since we are only interested in the low-energy physics of the putative symmetry-broken phase, constructing an appropriate IR field theory amounts to nothing more than constructing an appropriate kinetic term for $\phi$. 
To find a kinetic term invariant under a given multipole group $\mathcal{M}$, we need to find operators $D$ built out of spatial derivatives that annihilate all the polynomials in $\mathcal{M}$. If $a_\text{max}$ is the highest degree of the polynomials in $\mathcal{M}$, the simplest derivative operators which generically do the job are of the form 
\begin{align}
D = q^{I_{a_{\rm max}+1}} \p_{I_{a_{\rm max}+1}},
\end{align}
for some symmetric tensor $q^{I_{a_{\rm max}+1}}$. Although it is not generically possible to do so~\cite{Gromov2019}, we can sometimes find a set of $D_\alpha$ ($\alpha$ is an abstract index) with $s\le a_{\text{max}}$, where $a_\text{max}$ is the highest degree of the polynomials. In this case, the effective field theory will be invariant under some non-maximal multipole symmetry. 

Writing the invariant derivative operators as $D_\alpha$, the most general kinetic term is ~\cite{Gromov2019} 
\begin{align}
K[\phi(x)] &= g_{\alpha\beta} (D_\alpha \phi) (D_\beta \phi)
\end{align}
for some symmetric tensor $g_{\alpha\beta}$.
We will often write the Fourier transform of the kinetic term as $\phi_{-k} K_k \phi_k$.  Requiring some spatial symmetries restricts the choices of $g_{\alpha \beta}$. For the maximal multipole group, enforcing all rotation symmetries results in the kinetic term
\begin{align}
K_a[\phi(x)] &= (\partial_{I_{a+1}} \phi) (\partial_{I_{a+1}} \phi),
\end{align}
which is the kinetic terms studied in Ref.~\cite{Griffin2015}. 

For $a>0$, $K_a$ can be split into multiple terms while still remaining rotationally invariant. For example, the kinetic term for $\mathcal{M}^1_\max$ is 
\begin{align}
K_1[\phi(x)] = g_1 \big(\sum_{i} \partial_i \partial_i \phi\big)^2 + g_2 \sum_{i \ne j} (\partial_i \partial_j \phi)^2.
\end{align}
For the special cases of $g_2=0$ or $g_1=0$, the symmetry group expands to $\mathcal{M}^2_\text{sym}$ or $\mathcal{M}^2_\text{tr}$, respectively. However, for generic $g_1,g_2$ the symmetry group is $\mathcal{M}^1_\max$. Similar statements can be made for larger $a$.

Although we have been discussing field theories in the continuum, we will often care about theories that emerge from the low-energy degrees of freedom of a lattice model. In ordinary theories with monopole and lattice rotation symmetries, the lowest-order derivative term that can emerge from the lattice theory is $(\partial_i \phi)^2$, which is symmetric under the larger group of continuous rotations. Any terms that are not invariant under the continuous rotation group are irrelevant, so the continuous symmetry emerges. For multipole symmetries, leading-order terms may contain lattice anisotropies, so that the continuous rotation group does not emerge. For this reason we will only require that $\mmax{a}$ contain discrete rotation symmetries.

Finally, some of these multipole symmetries can be gauged to arrive at effective field theories for fracton phases. Of course, after gauging the theory will have gapless excitations due to the U(1) symmetry, but they can be Higgsed to arrive at a gapped phase. This is the sense in which the Haah group mentioned earlier corresponds to a field theory for Haah's code~\cite{BB}. See Ref.~\cite{BB, Gromov2019} for the full story. 

\section{Generalized Mermin-Wagner: full multipole breaking}\label{sec:full_breaking}

We are now ready to describe the generalized Mermin-Wagner argument in the case where an arbitrary maximal multipole group is spontaneously broken down to the trivial subgroup. This already appears in Ref.~\cite{Griffin2015}, so we are simply reviewing it here in preparation for the more generic cases to follow. Here, we will discuss both thermal and $T=0$ systems.  We will start with a heuristic argument in terms of domain wall energies before discussing a more careful diagnosis of spontaneous symmetry breaking in terms of correlation functions.

Consider first a clean system at $T>0$ with a spontaneously broken maximal multipole symmetry of degree $a$.
Then the kinetic term is proportional to 
\begin{align}
K[\phi(x)] = (\partial_{I_{a+1}} \phi)^2. \label{eqn:adisp}
\end{align} 
Symmetry breaking field configurations will be of the form $\phi(x) = a^0 + a^1_ix^i + \dots+ a^a_{I_a}x^{I_a}$, where $a^b$ are symmetric rank-$b$ tensors. This ensures there is no kinetic energy.  

Now consider nucleating domains of linear size $L$, in which the field takes on a different configuration $\phi(x) = c^0 + c^1_ix^i + \dots + c^a_{I_a}x^{I_a}$. The difference in the coefficients $c^b_{I_b}-a^b_{I_b}$ will be of order $L^{-b}$, as that is the natural scale introduced by the presence of the domain.
There is a $d$-dimensional region surrounding the domain, where the coefficients in $\phi(x)$ smoothly change their values over a length $L$. We will call this the thickened domain wall. 

Let us specialize to $d=1$ to simplify the analysis of the thickened domain wall. In higher dimensions, for large enough thickened domain walls, the coefficients of $\phi(x)$ will locally appear to only vary in one direction, so that the analysis reduces to the $d=1$ problem. In the thickened domain wall, we have 
\begin{align}
\phi(x) = b^0(y) + b^1(y) y + \dots + b^a(y) y^a,
\end{align}
where the $b^b(y)$ are polynomials of the dimensionless parameter $y=x/L$, with degrees constrained by the equations of motion coming from~\eqref{eqn:adisp}. Each derivative in~\eqref{eqn:adisp} pulls down a factor of $L^{-1}$, so the energy density will scale as $L^{-2(a+1)}$.

When we go back to $d$ dimensions, the energy density remains unchanged. We can integrate the density over the entire thickened domain wall to see that the total energy is $E\sim L^{d-2(a+1)}$.
In clean systems there is no energy gain from domain nucleation. When $d\le2(a+1)$, the energy cost is bounded, so the entropy gain favors domain creation. Thus, ordered phases are unstable and SSB cannot occur for $d\le2(a+1)$.

We can recover the result of this heuristic argument in a more formal way by considering correlation functions.
As a warm-up, consider the standard case where a monopole U(1) symmetry is spontaneously broken, leading to a Goldstone boson $\phi(x)$. The correlation function
\begin{align}
C(x) = \left\langle \phi(x) \phi(0) \right\rangle = \int \frac{d^dk}{(2\pi)^d} \frac{e^{i k \cdot x}}{k^2}, 
\end{align}
diverges when $d\le 2$. Correlation functions should not diverge, so the interpretation is that the Goldstone boson fluctuates strongly enough that the field $\phi(x)$ cannot be well defined. In turn, this tells us the symmetry could not have been broken. This is the classical Mermin-Wagner argument. We should note that, since we only care about long-distance divergences, we don't need to include the complex exponential as long as we only look for divergence at small $k$. We will thus drop this dependence in future calculations. 

When we consider a (maximal) multipole symmetry, the dispersion changes to $K_k=k^{2(a+1)}$. Now the correlation functions will scale as 
\begin{align}
C \sim \int \frac{d^dk}{K_k} \sim \int \frac{d^dk}{k^{2(a+1)}}, \label{eqn:correl}
\end{align}
which diverges for $d\le2(a+1)$. As before, we interpret the divergence of the correlation function as a sign that the symmetry cannot be spontaneously broken. This allows us to reproduce our scaling argument that SSB of the multipole group $\mathcal{M}^a_\max$ cannot occur for $d\le2(a+1)$.

In the zero temperature setting, the energy scaling no longer tells us where the critical dimension is. This is because even when there is no energy cost to forming domains, there is no entropy gain. Instead, quantum fluctuations must be the motivation for domain nucleation. To find the quantum critical dimension, we calculate the correlation function by way of the (imaginary-time) IR Lagrangian 
\begin{align}
     \mathcal{L}[\phi] = (\partial_\tau\phi)^2 + K[\phi].
\end{align} 
This gives 
\begin{align}
\label{tzerocorr}C(x) = \left\langle \phi(x) \phi(0) \right\rangle &\sim \int \frac{d\omega d^dk}{\omega^2 + k^{2( a + 1 )}},
\end{align}
which now diverges at $d\le a+1$~\cite{Griffin2015}, where $d$ is the number of spatial dimensions. We have halved the critical dimension, so that SSB cannot occur at $d\le a+1$. Note that the dynamical critical exponent in these theories is $z = a+1$, so that as expected, the classical and quantum critical dimensions are related by $d_\text{cl} = d_\text{q} + z$.

Furthermore, the structure of the quantum correlation function,
\begin{align}
C &\sim \int \frac{d\omega d^dk}{\omega^2 + K_k}\nn
&\sim \int \frac{d^dk}{\sqrt{K_k}},
\end{align}
suggests that, at least in some broad class of theories, the quantum critical dimension will be half the classical critical dimension. This is not true in general, though, as we will show.

\subsection{Non-maximal multipole group} \label{sub:nonmax}

We can also consider the fate of symmetry breaking for multipole groups other than the maximal multipole group. In general this will not match the critical dimension for the maximal group. 

For concreteness, we will consider some examples. First, recall the group $\mathcal{M}^2_\text{sym}$ from Sec.~\ref{sub:examples}. The group contains polynomials of degree 2. However, since the polynomials are all traceless, the most relevant derivative is $D = \sum_i \partial^2_i$.  We can immediately see the dispersion is $(k^2)^2$, so that the critical dimension is $d_\cl = 4$ or $d_\q = 2$.

We can also consider an anisotropic multipole group. 
Let the conserved charges be the monopole moment and $d_2$ components of the dipole moment. Then the dispersion will be $k^4 + p^2$, where $k$ has $d_2$ components, $p$ has $d_1$ components, and $d = d_1+d_2$. At nonzero temperature, the correlation function is
\begin{align}
C &\sim \int \frac{d^{d_2} k \, d^{d_1} p}{k^4 + p^2}\nn
&\sim \int x^{d_2/2 -1} p^{d_1-1} \frac{dx\, dp}{x^2 +p^2} \nn
&\sim \int q^{d_2/2 + d_1 - 2}\frac{dq}{q},
\end{align}
where $q^2 = x^2 + p^2 = k^4 + p^2$. 
If $d_1=2$ the symmetry can be broken for any $d_2>0$, and if $d_1=1$ then $d_2=2$  is critical. Of course, if $d_1=0$ we have an isotropic quartic dispersion and $d=d_2=4$ is critical. 

We can recover this result in  the energy-scaling argument. Since the system is anisotropic, the domains will have different sizes in different directions. Consider forming a domain of linear size $L_p$ in the  direction with quadratic dispersion and $L_k$ in the quartic direction. When the gradient of the order parameter coefficients is in the quadratic direction, the coefficients have to change over a length $L_p$ so the energy density is $L_p^{-2}$. Similarly, when the coefficients are changing in the quartic direction its energy density is $L_k^{-4}$. 

Both of these energy densities will be integrated over domain ``walls" with volume $L_k^{d_2} L_p^{d_1}$.
As a result, the total energy cost is
\begin{align}
E &\sim L_k^{d_2} L_p^{d_1-2} + L_k^{d_2-4} L_p^{d_1},
\end{align}
so that $L_p \sim L_k^2$ to make the terms match, and $E\sim L_p^{d_2/2+d_1-2}$. Recall that when $E$ does not increase for larger domains, the entropy gain will cause them to nucleate, destroying the ordered phase.
We see that the critical dimension $d= d_2 + d_1$ can be 2, 3, or 4, depending on how many directions have quadratic or quartic dispersions.

The $T=0$ correlation function behaves as
\begin{align}
C &\sim \int \frac{d\omega d^{d_2} k \, d^{d_1} p}{\omega^2 + k^4 + p^2}\nn
&\sim \int k^{d_2 - 2} p^{ (d_1-1)} \frac{k\,dk\, dp}{\sqrt{k^4 + p^2}} \nn
&\sim \int x^{d_2/2-1} p^{d_1-1} \frac{dx\, dp}{\sqrt{x^2 + p^2}} \nn
&\sim \int q^{d_2/2 -1} q^{d_1 -1} dq,
\end{align}
where again $q^2 = x^2 + p^2 = k^4 + p^2$. This diverges when $d_2/2 + d_1 \le 1$. 

The previous analysis suggests a procedure for finding the critical dimension in clean anisotropic systems for breaking the multipole symmetry to the trivial group. First, sort each dimension by the degree of its dispersion relation, so that there are $d_n$ dimensions with dispersion $k^{2n}$. Then define the effective dimension as 
\begin{align}
d_\eff = \sum_n \frac{d_n}{n}.
\end{align}
For $T>0$, we then conclude that SSB cannot occur if $d_\eff\le 2$, while at $T=0$ the critical dimension is $d_\eff = 1$.

We should emphasize that the statements in this subsection were framed in terms of the dispersion of the Goldstone modes rather than the structure of the symmetry group. It is always possible to find the dispersion given the multipole symmetry group, but it may require some basis changes (as in the Haah multipole group).

\section{Partial breaking of multipole symmetries} \label{sec:partial}

In the previous section we studied situations in which a multipole group was fully broken. In this section we turn our attention to examining what happens when the symmetry breaking is incomplete, with a nontrivial subgroup remaining unbroken. A simple example that illustrates why this problem is nontrivial is the following. 
 
Consider a situation in which a dipolar symmetry $\mcm^1_{\rm max}$ is spontaneously broken to its monopolar subgroup $\mcm^0_{\rm max}$, and specialize to the case of $d=3$ and $T>0$. In the spirit of Goldstone's theorem, the most naive thing to do when analyzing the symmetry broken phase would be to write down an action involving a set of $d=3$ Goldstone fields $\theta_j$, with $\t_j$ transforming linearly under the dipolar part of the symmetry group, but remaining invariant under the unbroken monopole subgroup, so that $\theta_j \mapsto \theta_j + \beta_j$ under a transformation parameterized by $P(x) = \alpha + \beta_j x^j$. In this case, the IR field theory for the putative symmetry-breaking phase would be of the form 
\begin{align}
\label{vector} \mathcal{L}= \sum_j  (\partial_\tau \theta_j)^2 + \sum_{i,j} g_{ij} (\partial_i \theta_j)^2.
\end{align}
As we are working in $d=3$ and at $T>0$, the theory \eqref{vector} indeed has SSB for the dipole symmetry, with $e^{i\t_j}$ developing long-range order. 

Of course, such a Goldstone field would break continuous rotation symmetries, but recall that we are only requiring that each maximal multipole group contain lattice rotation symmetries. We can then say that the theory~\eqref{vector} spontaneously breaks $\mmax{1}$ to $\mmax{0}$, in $d=3$ and at $T>0$.

The theory \eqref{vector} is not the only possibility consistent with this pattern of symmetry breaking, however. Indeed, consider instead the scalar field theory 
\begin{align}
\label{singlescalar}    \mathcal{L} = (\partial_\tau\phi)^2 + g_{ij} (\partial_i\partial_j\phi)^2, \end{align}
where $\phi$ transforms under the symmetry as $\phi \mapsto \phi + \alpha + \beta_jx^j$. 
 In the scalar theory \eqref{singlescalar}, the analysis of section \ref{sec:full_breaking} shows that the monopole symmetry is {\it not} spontaneously broken, with $e^{i\phi}$ possessing short-ranged correlation functions. However, the dipolar symmetry does indeed have SSB, and it is easy to check that the operators $e^{i\p_j\phi}$ --- which transform under the symmetry in the same way as $e^{i\t_j}$ --- have long-range order. Thus in this case, the two theories \eqref{vector} and \eqref{singlescalar} exhibit the same pattern of symmetry breaking, despite having a different number of Goldstone modes.\footnote{We will refer to the mode $\phi$ in the theory \eqref{singlescalar} as the ``Goldstone mode'' of the spontaneously broken dipole symmetry, since it is the only mode in the theory. Note however that it is $\p_j\phi$ (and not $\phi$ itself) which transforms under the symmetry action in the way expected of Goldstone bosons.} In order to formulate the correct version of the Mermin-Wagner theorem in the partial breaking case, it will be necessary to understand why this is so.

\subsection{Counting massless modes} \label{sub:subgroup}

That the number of Goldstone modes is not uniquely determined by the pattern of symmetry breaking is of course not in contradiction with Goldstone's theorem, as illustrated e.g. by the example of the ferromagnet vs the antiferromagnet. Indeed, correctly determining the number of Goldstone modes realized in a given situation depends not just on the symmetry breaking pattern, but also on information relating to how the various charge generators act on the ground state, as nicely described in \cite{watanabe2013redundancies}.

A general formal framework for understanding the counting of Goldstone modes in the present context can be developed by following the analysis of \cite{watanabe2013redundancies}, but here we will simply content ourselves with a heuristic physical argument. For simplicity, we will focus on systems in which the total symmetry group $\mcm_{\rm max}^a$ is spontaneously broken down to some nontrivial subgroup $\mcm_{\rm max}^b$. As mentioned previously, $\mmax{b}$ might only contain discrete rotation symmetries, so we will only concern ourselves with the spontaneous breaking of the polynomial generators. We will indicate that a generator corresponding to a polynomial of degree $c$ is spontaneously broken by saying that $\mmax{c}$ is broken.

Determining the counting of the Goldstone modes requires introducing the compressibilities $\kappa_c$, defined by 
\be \label{kappadef} \kappa_c^{I_c} = \frac{d\langle n_c^{I_c}\rangle}{d\mu_c^{I_c}},\ee 
where $n_c$ is the number density of $c$-pole charges, $\mu_c$ is a conjugate chemical potential, and $I_c = \{i_1,\dots,i_c\}$ is a composite index as before.

A necessary condition for $\mmax c$ to be spontaneously broken is that $\kappa_c \neq 0$. Indeed, if $\mmax c$ is spontaneously broken then $n_c$ creates Goldstones when acting on the ground state, thereby ensuring that the density-density correlator appearing in the calculation of \eqref{kappadef} must be nonzero at long wavelengths. Since we are assuming that $\mmax c$ is spontaneously broken for all $c>b$, we must accordingly have $\kappa_{c>b} \neq 0$. 

However, $\kappa_c \neq 0$ does not {\it necessarily} imply that $\mmax c$ is spontaneously broken: $\kappa_c\neq 0$ means that the system possesses gapless modes created by $n_c$, but it is possible for fluctuations to be strong enough so that these modes cannot be associated with the Goldstone modes of a spontaneously broken $\mmax c$ (this is of course what happens in the QLRO phase of the 1+1D XY model). Therefore, we are allowed to have $\kappa^c \neq 0$ even if $c \leq b$, without contradicting our assumption about the symmetry breaking pattern. 

Let then $s\leq a$ denote the smallest degree of multipole charge with nonzero compressibility, i.e. consider a situation in which $\kappa_c>0$ for all $s\leq c\leq a$, while $\kappa_c=0$ for all $c<s$. 
Given $s$, the minimal IR Lagrangian looks schematically like 
\be \label{generall} \mathcal{L} = \kappa_s (\partial_\tau \phi_{I_s})^2 + g_{J_{a+1}}(\partial_{J_{a-s+1}}\phi_{I_s})^2,\ee where 
$\p_{J_{a-s+1}} \equiv \p_{j_1} \cdots \p_{j_{a-s+1}}$, and where $\phi_{I_s}$ is a field that shifts by a polynomial of degree $a-s$ under the action of $\mmax a$.\footnote{The compressibility appears as the coefficient of the time derivative term since $\p_\tau \phi_{I_s}$ is the momentum conjugate to $\phi_{I_s}$ in the quantum theory, and hence correlation functions of fluctuations in $n_s$ are determined by those of $\p_\tau \phi_{I_s}$.}  If $s=b+1$, then $\phi_{I_s}$ may develop LRO while leaving $\mmax{b}$ invariant.

In order for $s< b+1$ to be consistent with our assumed pattern of symmetry breaking, it must be the case that the fluctuations of the fields in \eqref{generall} are strong enough so as to prevent $\mmax {s}$ from spontanoeusly breaking. The novelty provided by multipolar symmetries is that they offer an easy way of allowing for $s$ to be smaller than $b+1$, and thus provide an easy way of realizing multiple IR theories that have the same symmetry breaking pattern, but contain different numbers of modes. This is illustrated by the examples of the theories \eqref{vector} ($s=1$) and \eqref{singlescalar} ($s=0$), which have the same symmetry breaking pattern ($b=0$) in $d=3$ and at $T>0$. 

The theories we consider here have $(d+s-1 \text{ choose } s)$ massless modes, which is the number of degrees of freedom in a fully-symmetric rank-$s$ tensor. For the reasons discussed above, these modes may or may not be Goldstone modes. In different effective field theories, there may be other choices of symmetry-breaking fields and thus different numbers of independent modes. Still, we expect that in general the number of independent modes will depend on the compressibilities, not necessarily on the pattern of spontaneous symmetry breaking.

\subsection{Generalized Mermin-Wagner: partial multipole breaking} \label{sub:partial}

In the previous subsection, we saw that the number of Goldstone modes in the symmetry-breaking theory is determined by the smallest degree of multipole charge $s$ such that $\kappa_s\neq0$. We now turn to determining the symmetry breaking pattern that occurs for a given choice of $s$, spatial dimension $d,$ and temperature $T$, thereby allowing us to formulate a generalized version of the Mermin-Wagner theorem for the case of partial multipole breaking. 

Since $\kappa_c=0$ if $c<s$, none of the subgroups $\mmax {c<s}$ can be spontaneously broken. Suppose then that the subgroup $\mmax{c}$ is spontaneously broken, where $c\geq s$. To find the allowed values of $c$ consistent with this assumption, we need to calculate correlation functions of operators of the form $\exp(i\p_{J_{c-s}}\phi_{I_s})$, these being the operators transforming nontrivially under $\mmax c$ with the longest-range correlation functions. We have 
\be\ba \langle \p_{J_{c-s}}\phi_{I_s}(x) \p_{J_{c-s}}\phi_{I_s}(0)\rangle & \sim \int \frac{d\omega \, dk\, k^{d-1+2(c-s)}}{\omega^2 + k^{2(a-s+1)}} \\ 
& \sim \int dk\, k^{d-1+2(c-s) - \varsigma_T(a-s+1)}, \ea\ee 
where we have defined the function $\varsigma_T$ such that $\varsigma_T = 1$ if $T=0$ and $\varsigma_T = 2$ if $T>0$. Consequently, we find that $\mmax c$ can be spontaneously broken only in dimensions $d$ such that 
\be d>2(s-c) + \varsigma_T(a-s+1).\ee 
The preserved subgroup is then the largest multipole group $\mmax{b}$ that is not spontaneously broken.

For a given choice of symmetry breaking pattern (captured by $a$ and $b$) and a given choice of which compressibilities are nonzero (captured by $s$), the critical dimension for symmetry breaking is
\begin{align}
d_\text{lc} = 2(s - b-1) + \varsigma_T (a-s+1) \label{pbreak_mw} 
\end{align}
generalizing the Mermin-Wagner theorem to the partial breaking case. 
As expected, the lower critical dimensions in the quantum $(\varsigma_T = 1)$ and classical $(\varsigma_T = 2)$ cases are related by $d_{\rm cl} = d_{\rm q} + z$, where $z=a-s+1$ is the dynamical exponent of the Goldstone theory \eqref{generall}. 

If we only care about the pattern of symmetry breaking, $\mmax{a}\rightarrow \mmax{b}$, the critical dimension can be found by setting $s=0$, so that
\begin{align}
d_\text{lc} = \varsigma_T (a+1) -2 (b+1). \label{pbreak_mw_nos}
\end{align}
As a sanity check on \eqref{pbreak_mw_nos}, note that for full multipole breaking, $b=-1$ (recall that $\mmax{-1}$ contains only translations and rotations), we recover the lower critical dimension of $\varsigma_T(a+1)$ obtained in the previous section. 
In any higher dimension, such a symmetry breaking pattern will be possible. However, $s$ must always be chosen so as to give the correct preserved subgroup. As emphasized earlier, the number of Goldstone modes depends on $s$, not $b$.

As an interesting example, consider the case where the dipolar group $(a=1)$ is broken down to the monopole group $(b=0)$. At $T=0$, we see from \eqref{pbreak_mw_nos} that this is possible provided that $d>0$. Thus the partial breaking of dipole symmetry provides us with an example where a continuous symmetry can be spontaneously broken in $d=1$ at zero temperature. 

\section{Systems with quenched disorder} \label{sec:disord}

Let us now add some quenched disorder to our systems. In particular, we will consider disorder that explicitly breaks the symmetry locally but does not break the symmetry on average. Spatial disorder will also break translation and rotation symmetry, but again not on average. Of course, strong enough disorder can always destabilize the ordered phase, so will not consider that case. Weak disorder can discourage the ordered phase and raise the critical dimension at which SSB is impossible. Theorems of this type originate with Imry and Ma~\cite{ImryMa} and were proved by Aizenman and Wehr~\cite{Aizenman}.

In a disordered classical system, the energy scaling argument gives a nice explanation for why the critical dimension changes. Consider the formation of a domain of linear size $L$ in an otherwise ordered phase fully breaking the maximal multipole group $\mathcal{M}^a_\max$. There are $\sim L^d$ disorder samples in the new domain. Since each sample is taken independently, the typical energy gain from forming the domain will be $\sim L^{d/2}$, by the central limit theorem. Comparing this to the cost of domain formation we calculated previously, the ordered phase will be unstable to domain nucleation when $d\le 4(a+1)$. We emphasize that we are here considering disorder that couples linearly to the order parameter. 

In clean systems, the transition from classical to quantum brought down the critical dimension because the argument depended on entropy. The Imry-Ma argument instead only appeals to energy considerations, so the quantum critical dimension in the presence of disorder remains the same as the classical critical dimension~\cite{Vojta2013}. In disordered quantum systems, full SSB is impossible for $d\le 4(a+1)$.

To reproduce this argument using a correlation function calculation~\cite{ImryMa} we need to calculate the response to disorder. Call the disorder field $h(x)$ and redefine $\phi(x)$ to be the variation away from the average value $\phi_0$. The relevant part of the Hamiltonian is 
\begin{align}
H &\sim \int d^dx \left[ \half K[ \phi(x)] - \phi(x) h(x) \right] \nn
&\sim \int d^dk \left[ \half \phi_{-k} K_k \phi_k - \phi_{-k} h_k \right] \nn
&\sim \int d^dk \left[ \half \phi_{-k}' K_k \phi_k' - \half h_{-k} K^{-1}_k h_k \right],
\end{align}
where we have written $\phi_k' = \phi_{-k} - K^{-1}_k h_{-k}$. Then, from the expectation of $\phi_k$,
\begin{align}
\langle \phi_k \rangle &\sim \frac{\delta Z}{\delta h_{-k} } \nn
&\sim K_k^{-1} h_{k},
\end{align}
we can see that the disorder produces fluctuations mediated by the susceptibility.

We can then compute the correlation function for $\phi(x)$,
\begin{align}
\langle \phi(0) \phi(x) \rangle &\sim \int d^dk \, K_k^{-2} \langle h_{-k} h_{k} \rangle e^{ik\cdot x},
\end{align}
where $\langle h_{-k} h_{k} \rangle e^{ik\cdot x}$ does not affect the divergence at small $k$, assuming the disorder is short range correlated. We can compare to~\eqref{eqn:correl} to see that in the presence of disorder that couples linearly to an order parameter fully breaking multipole symmetry, the critical dimension for a disordered system is twice the critical dimension for having direct full multipole symmetry breaking in a clean classical system.

We will now consider disorder that breaks the symmetry to a subgroup. For simplicity, let the symmetry of the system be the maximal dipole group and let the disorder break the dipole part of the group but not the monopole part. The kinetic term is $K[\phi(x)]= (\partial_i \partial_j \phi)^2$, while the coupling to disorder will be something like $(\partial_i\phi - \zeta_i)^2$, where $\zeta_i$ is the disorder field. 
Define the vector field $\xi_i = \partial_i\phi$, with kinetic term $K[\xi(x)]= (\partial_i \xi_j)^2$. In the ordered phase $\xi_i$ will have some constant value. It need not vanish because of the dipole symmetry. In the presence of weak disorder, it will want to follow the $\zeta_i$ field where possible. We can now follow the argument of the Imry-Ma theorem to say that the critical dimension will be $d=4$.

A more general case is a system with symmetry $G = \mathcal{M}_\max^a$ and disorder that breaks the symmetry to $\mathcal{M}_\max^b$ but preserves $G$ on average. In this system the critical dimension for long-range order of a field that breaks $\mmax{a}$ but preserves $\mmax{b}$ should be $d=4(a-b)$ for both classical and quantum systems. Again, let $b=-1$ denote the trivial group in order to recover our previous results.

\section{Explicit lattice example} \label{sec:example}

We now present a transparent quantum lattice example to illustrate the spontaneous breaking of multipolar symmetry. We will not derive further results from this model. Rather, we apply our previous results to this model to understand how various phases are available in different dimensions. Reference~\cite{dbhm} gives detailed descriptions of the various phases and their transitions.

Consider a $d$-dimensional hypercubic lattice with sites labeled by $j$ and directions labeled by $\mu = 1,\dots,d$. Let there be bosonic degrees of freedom annihilated by $b_j$ on the sites, and `dipolar' bosonic degrees of freedom annihilated by $d_{j,j+\mu}$ on the edges. We can thus think of there as being $d$ species of edge-type bosons.

Define the symmetry operators
\begin{align}
\op{1}(\xi) &= \prod_je^{i\xi n_j}\nn
\op{2}_\mu(\xi) &= \prod_j e^{i\xi (j_\mu n_j)}e^{i\xi n^d_{j,j+\mu}},
\end{align}
where $j_\mu$ is the $\mu$-th component of the site label $j$, $n$ is the site boson number operator, and $n^d$ is the edge boson number operator.
The first operator corresponds to conservation of the total number of site bosons, while the $\mu$-th component of the second operator corresponds to the sum of the $\mu$-th component of the total dipole moment of site bosons and the total number of the $\mu$-th type of edge bosons. These symmetries allow us to exchange a dipole of site bosons for an edge boson. 

The minimal Hamiltonian obeying these symmetries is  
\begin{align}
H_0  &= H_b + H_d + H_\text{int}\nn
H_b &= \sum_j \left[ \frac{U_b}{2}  n_j (n_j-1) - \mu_b \, n_j \right] \nn
&\qquad - \sum_{j,\mu,\nu} t_b \left[  b_j b^{\dag}_{j+\mu} b^{\dag}_{j+\nu} b_{j+\mu+\nu} + H.C. \right]\nn
H_d &= \sum_{j,\mu} \left[ \frac{U_d}{2} n^d_{j,j+\mu} (n^d_{j,j+\mu} - 1) -\mu_d \, n^d_{j, j + \mu} \right]\nn
&\qquad  - \sum_{j,\mu,\nu} t_d \left[d^{\dag}_{j, j + \mu} d_{j + \nu, j+\mu + \nu} +H.C.\right]\nn
H_\text{int} &= \sum_{j,\mu} g\left[ b_j d_{j,j+\mu} b^\dag_{j+\mu} + H.C. \right],
\end{align}
where sums are taken over sites $j$ and lattice directions $\mu$ and $\nu$.  
For simplicity, let us set 
\begin{align}
\frac{\mu_b}{ U_b} = \frac{\mu_d }{ U_d} = \frac{1}{2},
\end{align}
so there is a robust insulating phase at weak hopping $t_b, t_d$~\cite{Fisheretal}. A detailed discussion of the phase diagram and phenomenology of this model is given in~\cite{dbhm}; in the following we will simply discuss the symmetry breaking patterns realized in various different limits. 

The Hamiltonian $H_b$ controls condensation of the site bosons. When $t_b$ is small, the site bosons are in number eigenstates while when $t_b$ is large the site bosons condense (although note that since the term proportional to $t_b$ is quartic in the site boson operators, the condensation transition may be nonstandard). Similarly, the edge bosons condense when $t_d$ is large. Note that lattice rotation symmetry will be broken if only a subset of the edge bosons condense, while if all $d$ species of condense equally then lattice rotation symmetry is preserved. We are not aware of a way to condense objects with dipolar charge while preserving {\it continuous} rotation symmetry. 

The term $H_\text{int}$ is allowed by the symmetry because it simultaneously removes an edge boson and creates a dipole of site bosons. Note that when the edge bosons are condensed, $H_\text{int}$ gives the site bosons an effective single particle hopping: thus condensation of single bosons can either proceed directly (by making $t_b$ large) or indirectly (by first condensing dipoles, and making the effective single-particle hopping term large). 

Consider first the phase in which only the edge bosons have condensed, with the site bosons remaining gapped. In this case the dipole compressibility $\kappa_1$ is nonzero, while the charge compressibility vanishes, $\kappa_0=0$. The condensed phase is therefore described in the IR by the $z=1$ theory \eqref{vector}, and the dipolar symmetry is spontaneously broken down to the monopole subgroup provided that $d>1$ (for $T=0$) or $d>2$ (for $T>0$). 

Next, consider the phase where the site bosons have condensed. In this case both $\kappa_1,\kappa_0$ are nonzero, and deep in the condensed phase the IR physics is described by the $z=2$ theory \eqref{singlescalar}. The dipolar symmetry is spontaneously broken in $d>0$ if $T=0$ and in $d>2$ at $T>0$, while the monopole subgroup is spontaneously broken in $d>2$ at $T=0$ and $d>4$ at $T>0$. This model therefore provides us with a way of realizing symmetry-broken phases described by both of our earlier example theories \eqref{vector} and \eqref{singlescalar}.

Let us now consider the same system, but with disorder.  The disorder Hamiltonian is
\begin{align}
H_\text{dis} &= \left[h_b\sum_j \sigma^*_jb_j + h_d \sum_{j,\mu} \sigma^*_{j,\mu}d_{j,j+\mu}\right] + H.C.,
\end{align}
where $h_b$ and $h_d$ control the magnitude of the disorder and each instance of $\sigma$ is a random phase. We will always consider $h_b,h_d\ll1$.

For $h_b,h_d\ne 0$, we can fully rely on the Imry-Ma argument. No symmetry breaking can occur for $d\le 4$, the dipole part of the symmetry may be broken for $4<d\le 8$, and any symmetry-breaking phase can occur for $d>8$. 

\section{Discussion} \label{sec:disc}

In this paper we analyzed the spontaneous symmetry breaking of various multipole groups and discussed generalized Mermin-Wagner theorems for breaking a maximal multipole group, either fully or to a subgroup. We also considered multipole groups that are not the maximal multipole group, and the effect of quenched disorder. The disorder that we considered explicitly broke the symmetry, either fully or to a subgroup.

Of course, we could consider further combinations of effects. For example, we could spontaneously break a symmetry from a group $G$ to a subgroup $H$, where one or both of $G,H$ are non-maximal; we could also consider non-maximal groups with disorder. While the number of potential examples to consider is large, they should all be analyzable using the ideas introduced herein. 

We should be clear that the arguments in this paper are Mermin-Wagner or Imry-Ma arguments. 
In principle, even when the Gaussian symmetry breaking fixed point is unstable to fluctuations, a non-trivial fixed point with long range order could arise (see e.g. \cite{TonerRadzihovsky}). Whether and when such non-trivial fixed points can be realized in models with multipolar symmetry is an important problem for future work. 

We also emphasize that when the ordered phase does not exist, we did not provide any argument for what phase should replace it. For example, in 2 spatial dimensions there can be no ordered phases of continuous monopole (ordinary) symmetries. For $O(n)$ models with $n>2$, the result is that the disordered phase is the only phase \cite{polyakov}. For $n=2$, there can in addition be quasi-long-range-ordered phases. Determining what kind of phase {\it can} be obtained in the absence of long range ordered symmetry breaking would at a minimum require understanding the nature and role of topological defects in the symmetry breaking order parameter, akin to vortices in the XY model. We expect that if we stick with an $O(2)$ internal group, then quasi-long range order should always be a possibility at the marginal spatial dimension, at least in the limit where the vortex core energy is infinitely large. The extension to non-Abelian internal groups is beyond the scope of this work. 
The range of possible symmetry unbroken phases could also be even richer in the presence of disorder, where various glassy phases could also come into play \cite{Fisheretal}. Our discussion of disorder physics was also limited to quenched short-range correlated disorder. Extensions to disorder with long-range correlations, or annealed disorder, are left to future work.

We should also emphasize that our discussion has utilized standard concepts from statistical physics, which in turn amounts to assuming ergodicity. However, quantum dynamics with multipolar symmetries can break ergodicity \cite{KHN, Sala}, in which case our analysis would not straightforwardly apply. It is however believed that the strict ergodicity breaking is limited to systems with strictly short range interactions (below some critical range) and that systems in which the interactions have long range tails (whether power law or exponential) should generically obey ergodicity at long times (although see \cite{NS}). Since long range tails are generic in physical systems, we believe our arguments should generically apply.  

Another setting for generalized Mermin-Wagner-type arguments is higher form global symmetries~\cite{GKSW, Lake, Marvin}. It would also be interesting to see what sort of subtleties could exist in the spontaneous breaking of those symmetries, through partial symmetry breaking or disorder. Since the order parameters for higher-form symmetries are nonlocal, it is difficult to couple disorder directly to the order parameters. In the case of arbitrary perturbations the symmetry becomes broken microscopically, but emerges at long wavelengths. Could disorder have any effect on the Mermin-Wagner behavior of higher-form symmetries? We leave these questions for future work. 

Finally, there exists a body of work on generalized Mermin-Wagner arguments in systems with subsystem symmetries \cite{Batista2005, SeibergA, SeibergB, SeibergC, Gorantla2021, Distler2021}. Subsystem symmetries are rather different in character to the multipolar symmetries discussed herein, but are also related to fracton phases via duality \cite{VHF2}. There can be theories with subsystem symmetries where symmetry breaking cannot occur even above the critical dimension, due to the UV/IR mixing~\cite{Gorantla2021}. However, it is always possible to write down theories that saturate the generalized Mermin-Wagner bound~\cite{Distler2021}.
Exploration of connections between the present work and the literature on subsystem symmetries would also be a fruitful topic for future work. 

{\bf Acknowledgements} We thank Leo Radzihovsky for pointing out that Mermin-Wagner instability could mean flow to ordered but non-trivial fixed points. This work was supported by
the U.S. Department of Energy, Office of Science, Basic Energy Sciences, under Award \# DE-SC0021346. E.L. was supported by a Hertz Fellowship.

\end{document}